\documentclass[10pt,twocolumn]{article} 

\usepackage{amsmath}
\usepackage{graphicx}
\usepackage{amssymb}
\begin{document}

\title{\bf
Gravitating Cho-Maison Monopole} 

\author{{Khai-Ming Wong, Dan Zhu and Guo-Quan Wong}\\
\textit{{\small School of Physics, Universiti Sains Malaysia, 11800 USM, Penang, Malaysia}}}

\date{March 2021}
\maketitle

\begin{abstract}
We study numerical solutions corresponding to spherically symmetric gravitating electroweak monopole and magnetically charged black holes of the Einstein-Weinberg-Salam theory. The gravitating electroweak monopole solutions are quite identical to the gravitating monopole solution in SU(2) Einsten-Yang-Mills-Higgs theory, but with distinctive characteristics. We also found solutions representing radially excited monopole, which has no counterpart in flat space. Both of these solutions exist up to a maximal gravitational coupling before they cease to exist. Lastly we also report on magnetically charged non-Abelian black holes solutions that is closely related to the regular monopole solutions, which represents counterexample to the `no-hair' conjecture.
\end{abstract}

\section{Introduction}
\label{sec:1}

Since the introduction of Dirac monopole by P.A.M. Dirac \cite{kn:1}, magnetic monopole has become an subject that attracts a lot of interest, both theoretically and experimentally. Since then the Dirac monopole has been generalized to non-Abelian monopoles, most notably Wu-Yang monopole in SU(2) Yang-Mills theory \cite{kn:2} and 't Hooft-Polyakov monopole in SU(2) Yang-Mills-Higgs (YMH) theory \cite{kn:3}. While the Dirac monopole and Wu-Yang monopole possess infinite energy due to the presence of point singularity in the solutions, the 't Hooft-Polyakov monopole possesses finite energy with no singularity found anywhere. The mass of 't Hooft-Polyakov monopole was estimated to be of order 137 $M_{\scalebox{.6}{\mbox{W}}}$, where $M_{\scalebox{.6}{\mbox{W}}}$ is the mass of intermediate vector boson.

The coupling of gravity to the SU(2) YMH theory, known as the SU(2) Einstein-Yang-Mills-Higgs (EYMH) theory has been shown to possess important solutions \cite{kn:4}. These solutions include globally regular gravitating monopole solutions, radial excitation and magnetically charged black hole solutions. For small gravitational coupling, the gravitating monopole solution emerges smoothly from flat space 't Hooft-Polyakov monopole. The (normalized) mass of gravitating monopole solution decreases with increasing gravitational coupling and the solution ceases to exist beyond a maximal value of gravitational coupling. Besides the fundamental gravitating monopole there exists radially excited monopole solution, where the gauge field function of the $n$-th excited monopole possess $n$ nodes, and this is different from the gauge field function of fundamental monopole solution that decreases monotonically to zero. Having no flat space counterparts, these excited solutions are related to the globaly regular Bartnik-Mckinnon solutions in SU(2) Einstein-Yang-Millls (EYM) theory \cite{kn:5}. There also exist magnetically charged EYMH black hole solutions which represent counterexamples to the `no-hair' conjecture. Distinct from the embedded Reissner-Nordstrom (RN) black holes with unit magnetic charge, these black hole solutions emerge from the regular magnetic monopole solutions when a finite regular event horizon is imposed. Consequently, they have been characterized as `black holes within magnetic monopoles'.
 
The SU(2) $\times$ U(1) Weinberg-Salam theory has been shown to possess important topological magnetic mono-pole solution, known as the electroweak monopole or simply Cho-Maison monopole \cite{kn:6}. As a hybrid between Dirac monopole and 't Hooft-Polyakov monopole, the Cho-Maison monopole describes a real monopole dressed by the physical W-boson and Higgs field. Although the Cho-Maison monopole has a singularity at the origin which makes the energy divergent, it has been shown there are ways to regularize the energy and estimating the mass at 4 to 10 TeV \cite{kn:7,kn:8,kn:9}. Recently, there is also reports on a more natural way to regularize the energy, suggesting that the new BPS bound for the Cho-Maison monopole may not be smaller than 2.98 TeV, more probably 3.75 TeV \cite{kn:10}. As mentioned in Refs. \cite{kn:7,kn:8,kn:9,kn:10}, the non-triviality of electromagnetic U(1) ensures that the electroweak monopole must exist and this makes the experimental detection of electroweak monopole an urgent issue after the discovery of Higgs boson. For this reason experimental detectors around the globe are actively searching for magnetic monopole \cite{kn:11,kn:12,kn:13,kn:14}. 

Recently gravitationally coupled electroweak mono-pole solutions in Einstein-Weinberg-Salam (EWS) theory has also been reported by Cho et al. \cite{kn:15}. Their results confirm the existence of globally regular gravitating electroweak monopole solution, before changes to the magnetically charged black hole as the Higgs vacuum value approaches to the Planck scale.

In this paper, we study in more detail the gravitating electroweak monopole in EWS theory, and report additional radially excited electroweak monopole solution, as well as the corresponding `black hole within electroweak monopole' solutions of the EWS theory. Our results therefore confirm that all solutions found in the SU(2) EYMH theory \cite{kn:4} have their corresponding counterpart in the EWS theory, but with distintive functional behaviour. From the physical point of view, these solutions are very important as Weinberg-Salam theory itself is a realistic theory.   

\section{Einstein-Weinberg-Salam Theory}
\label{sec:2}

We consider the SU(2)$\times$ U(1) EWS action as
\begin{equation}
S = S_G +  S_M = \int L_G \sqrt{-g}~d^4x + \int L_M \sqrt{-g}~d^4x,
\label{eq.1}
\end{equation}
with 
\begin{equation}
L_G = \frac{R}{16 \pi G} ,
\label{eq.2}
\end{equation}
and 
\begin{eqnarray}
L_M &=& - \frac{1}{4}F^a_{\mu\nu} F^{a\mu\nu} - \frac{\epsilon \left( \phi \right)}{4} f_{\mu\nu} f^{\mu\nu}  \nonumber\\
&-&  \left( \hat{D}_\mu \phi \right)^{\dagger}  \left( \hat{D}^\mu \phi \right) - \frac{\lambda}{2} \left( \phi^{\dagger} \phi- \frac{\mu^2}{\lambda} \right)^2,
\label{eq.3}
\end{eqnarray}
where
\begin{equation}
\hat{D}_{\mu} \phi = \left( \partial_{\mu} - \frac{ig}{2} \sigma^a A^a_{\mu} - \frac{ig'}{2} B_{\mu} \right)  \phi,
\label{eq.4}
\end{equation}
in which $\hat{D}_{\mu}$ is the covariant derivative of the SU(2) $\times$ U(1) group. The function $\epsilon \left( \phi \right)$ in Eq.(\ref{eq.3}) is a positive dimensionless function of the Higgs doublet which tends to unity asymptotically. In general, $\epsilon \left( \phi \right) $ modifies the permeability of the hypercharge U(1) gauge field while retaining the SU(2) x U(1) gauge symmetry \cite{kn:15}.

To construct globally regular gravitating monopole and magnetically charged black hole, we consider the spherically symmetric Schwarzschild-like metric
\begin{equation}
ds^2 = - N^2 A ~dt^2 + \frac{1}{A} dr^2 + r^2 d\theta^2 + r^2 \sin^2\theta d\phi^2,
\label{eq.5}
\end{equation}
with
\begin{equation}
A = 1- \frac{2G  m}{r},
\label{eq.6}
\end{equation}
and the following electrically neutral ansatz for the matter functions,
\begin{eqnarray}
&& \phi = \frac{H}{\sqrt{2}} ~\xi,~~~\xi = i \begin{bmatrix}
    \frac{\sin\theta}{2} e^{- i \phi} \\
    -\frac{\cos\theta}{2} 
  \end{bmatrix},   \nonumber\\
&&  A^a_0 = 0,~~ A^a_{i} = - \frac{\left( 1 - K \right) }{g r} \hat{\phi}^a \hat{\theta}_i + \frac{\left( 1 - K \right) }{g r} \hat{\theta}^a \hat{\phi}_i,  \nonumber\\
&& B_0 = 0,~~~ B_i = - \frac{1}{g'} \frac{\left( 1 - \cos\theta \right)}{r \sin\theta} \hat{\phi}_i. 
\label{eq.7}
\end{eqnarray}
Here $N$, $A$, $m$, $K$ and $H$ are all functions of $r$. 

The $tt$ and $rr$ components of the Einstein equations then yield the equations for the metric functions 
\begin{equation}
\frac{N'}{N} = 4 \pi G r \left(  \frac{2  K'^2 }{g^2 r^2}+ H'^2  \right),
\label{eq.8}
\end{equation}
and 
\begin{eqnarray}
m' &=& 4 \pi r^2   \left\{ A \left(  \frac{K'^2}{g^2 r^2} + \frac{H'^2}{2} \right)  + \frac{\left( K^2 - 1 \right)^2}{2 g^2 r^4}  \right. \nonumber\\ 
&+&   \left. \frac{\lambda}{8} \left( H^2 - \frac{2 \mu^2}{\lambda} \right)^2 + \frac{\epsilon}{2 g'^2 r^4} + \frac{1}{4} \frac{H^2 K^2}{r^2} \right\}.
\label{eq.9}
\end{eqnarray}
The equations for the matter functions read
\begin{equation}
A K'' + \left(  A' + A \frac{N'}{N} \right) K' +\frac{ \left( 1-K^2 \right) K }{r^2} -\frac{1}{4} g^2 H^2 K = 0,
\label{eq.10}
\end{equation}
and 
\begin{eqnarray}
&& A H'' + \left(  A' + \frac{2 A}{r} + A \frac{N'}{N} \right) H' - \frac{H K^2}{2 r^2}    \nonumber\\
&& - \frac{\lambda}{2} \left( H^2 - \frac{2 \mu^2}{\lambda} \right) H - \frac{1}{2 g'^2 r^4 } \frac{d \epsilon \left( H \right)}{dH} = 0.
\label{eq.11}
\end{eqnarray} 
Prime denotes derivative with respect to $r$, and $H_0 = \sqrt{2} \mu/\sqrt{\lambda}$ is the Higgs vacuum expectation value.

To facilitate numerical calculation, we consider the following dimensionless coordinate $x$ and dimensionless mass function $\widetilde{m}$,  
\begin{equation}
x = M_{\scalebox{.5}{\mbox{W}}} r,~~~ \widetilde{m} = G M_{\scalebox{.5}{\mbox{W}}} m,
\label{eq.12}
\end{equation} 
with $M_{\scalebox{.5}{\mbox{W}}} = \frac{1}{2} g H_0$. The Higgs field is also rescaled as $H \rightarrow H_0 H$ and the solutions then depend on coupling constant $\alpha$ and $\beta$, where
\begin{equation}
\alpha^2 = 4 \pi G H_0^2,~~~\beta^2 = \frac{\lambda}{g^2},
\label{eq.13}
\end{equation} 
as well as the Weinberg angle $\theta_{\scalebox{.5}{\mbox{W}}}$. 

With Eqs.(\ref{eq.12})-(\ref{eq.13}), the full set of Eqs.(\ref{eq.8})-(\ref{eq.11}) transform into
\begin{eqnarray}
&& \frac{1}{N} \frac{dN}{dx} = \alpha^2 x \left[  \frac{1}{2 x^2} \left(  \frac{d K}{dx} \right)^2 + \left(  \frac{d H}{dx} \right)^2  \right], \nonumber
\end{eqnarray}

\begin{eqnarray}
&& \frac{d \widetilde{m}}{dx}= \alpha^2 x^2   \left\{ \frac{A}{2} \left[ \frac{1}{2 x^2} \left(  \frac{d K}{dx} \right)^2 + \left(  \frac{d H}{dx} \right)^2 \right]  \right. \nonumber\\
&& \left. +  \frac{\left( K^2 - 1 \right)^2}{8 x^4} + \frac{\beta^2}{2} \left( H^2 - 1 \right)^2 + \frac{\epsilon}{8 \omega^2 x^4} + \frac{H^2 K^2}{4 x^2} \right\}, \nonumber
\end{eqnarray}

\begin{eqnarray}
&& A \frac{d^2 K}{dx^2} + \left(  \frac{dA}{dx} + \frac{A}{N} \frac{dN}{dx} \right) \frac{dK}{dx} \nonumber\\
&& + \frac{ \left( 1-K^2 \right) K }{x^2} -  H^2 K = 0,  \nonumber
\end{eqnarray}

\begin{eqnarray}
&& A \frac{d^2 H}{dx^2} + \left(  \frac{dA}{dx} + \frac{2 A}{x} + \frac{A}{N} \frac{dN}{dx} \right) \frac{dH}{dx} - \frac{H K^2}{2 x^2} \nonumber\\
&&  - 2 \beta^2 \left( H^2 - 1 \right) H - \frac{1}{2 \omega^2 x^4} \frac{d \epsilon}{d H}= 0, 
\label{eq.14}
\end{eqnarray}
where $\omega = g'/g = \tan \theta_{\scalebox{.5}{\mbox{W}}}$. Here we consider physical value of $\omega = 0.53574546$ by adopting $\sin^2\theta_{\scalebox{.5}{\mbox{W}}} = 0.22301323$  \cite{kn:16}. Since $ M_{\scalebox{.5}{\mbox{H}}} = \sqrt{2} \mu$ and $M_{\scalebox{.5}{\mbox{W}}} = \frac{1}{2} g H_0$, we may also put Eq. (\ref{eq.13}) in the form of 
\begin{eqnarray}
\alpha = \sqrt{ 4 \pi G} H_0 = \sqrt{4 \pi} \frac{H_0}{M_{\scalebox{.5}{\mbox{P}}} },~~~\beta = \frac{1}{2} \frac{M_{\scalebox{.5}{\mbox{H}}} }{M_{\scalebox{.5}{\mbox{W}}} },
\label{eq.15}
\end{eqnarray}
where by adopting physical values of $M_{\scalebox{.5}{\mbox{H}}} = 125.10$ GeV and $M_{\scalebox{.5}{\mbox{W}}}$ = 80.379 GeV, the physical value of $\beta$ used here is 0.77818833.

Obviously solutions to Eqs.(\ref{eq.14}) depend on the permeability function $\epsilon$. In the paper by Cho et al. \cite{kn:15}, the form of $\epsilon = \left(  H / H_0 \right)^8$ is considered. In a recent paper \cite{kn:10}, $\epsilon = \left(  H / H_0 \right)^n$ is also possible. For the sake of simplicity, we considered $n = 8$ in this paper. However we would like to point out that all values of $n = 1, 2, 3, ...8$ seem to produce convergent numerical results. 

As our solution is electrically neutral, following Ref. \cite{kn:17}, we consider a special solutions of Eqs.(\ref{eq.14}), which is the embedded RN solutions with mass $\widetilde{m}_{\infty}$ and magnetic charge near unity, 
\begin{eqnarray}
&& \widetilde{m}(x) = \widetilde{m}_{\infty} - \frac{\alpha^2 }{8 x} \left( 1 + \frac{\epsilon}{\omega^2}  \right) ,~~ N(x) = 1, \nonumber\\
&&  K(x) = 0,~~ H(x) = 1,
\label{eq.16}
\end{eqnarray} 
where we will consider $\epsilon = 1$ as $H = 1$. The corresponding extremal RN solutions then possess horizon $x_{\scalebox{.5}{\mbox{H}}}$, where
\begin{eqnarray}
x_{\scalebox{.5}{\mbox{H}}} = \widetilde{m}_{\infty} = \frac{\alpha}{2} \sqrt{1+\frac{1}{\omega^2}}.
\label{eq.17}
\end{eqnarray} 
From Eq.(\ref{eq.12}) and Eq.(\ref{eq.16}), the ADM mass can be defined as
\begin{eqnarray}
m_{\scalebox{.5}{\mbox{ADM}}}  = \frac{4 \pi H^2_0 }{M_{\scalebox{.5}{\mbox{W}}}}  \frac{\widetilde{m}_{\infty}}{\alpha^2},
\label{eq.18}
\end{eqnarray} 
where one can readily read off the value for ADM mass from the plot of $\widetilde{m}_{\infty}/\alpha^2$ versus $\alpha$.

\section{Gravitating Monopole}
\label{sec:3}

\begin{figure}[!b]
	\centering
	\hskip0in
	 \includegraphics[width=3.3in]{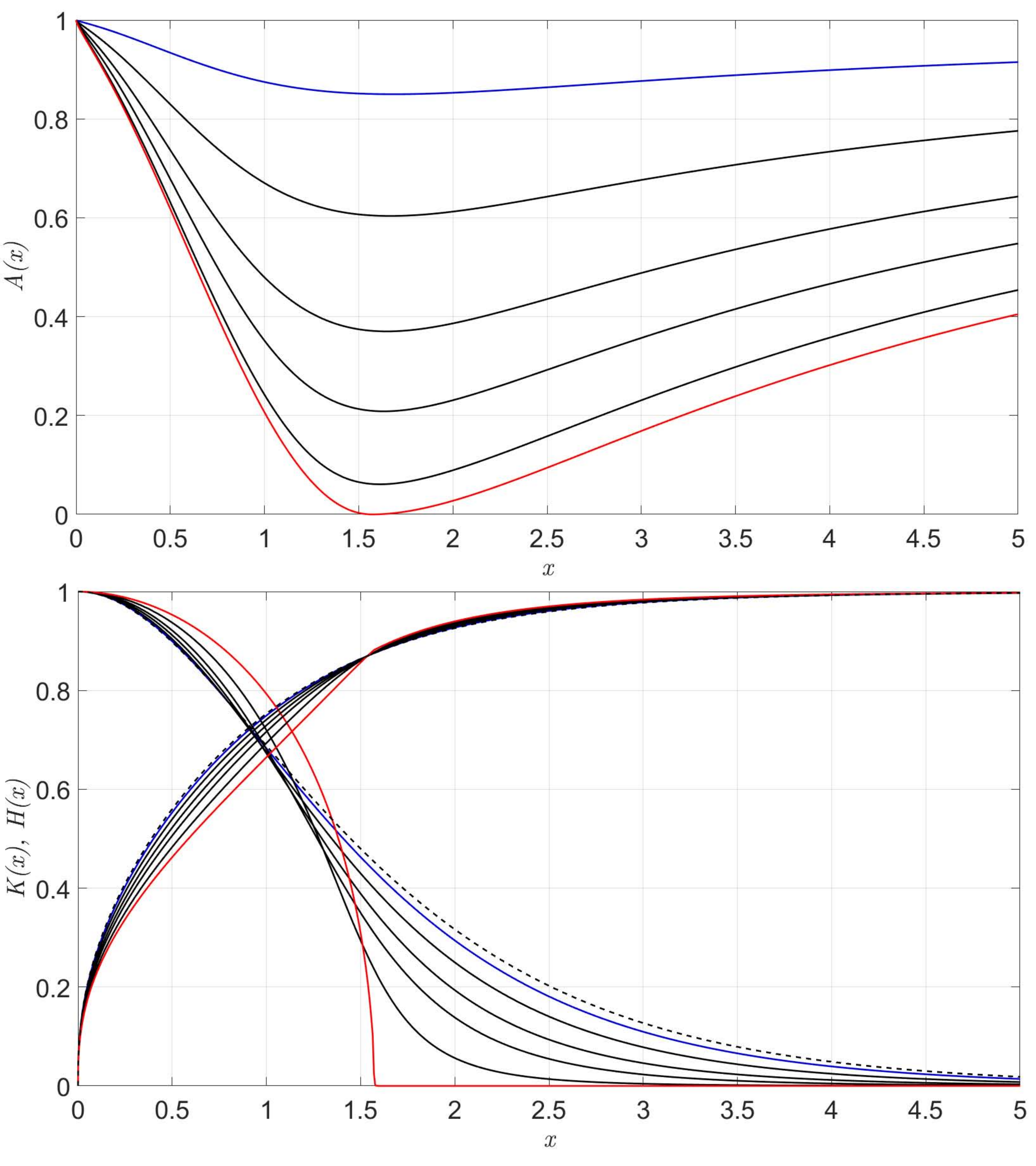} 
	\caption{Functions of $A(x)$, $K(x)$ and $H(x)$ versus $x$ of the fundamental gravitating electroweak monopole with physical $\beta$ and Weinberg angle for $\alpha =0.6 $ (blue), $1.0, 1.3, 1.5, 1.7$ and $1.814$ (red). Dashed line indicates non-gravitating monopole. }
 \label{Fig.1}
\end{figure}

We first consider globally regular gravitating monopole solutions. Asymptotic flatness requires that the metric functions $N$ and $A$ both approach a constant at spatial infinity. We here adopt
\begin{eqnarray}
N(\infty) = 1, ~~~\widetilde{m}(\infty) = \widetilde{m}_{\infty}.
\label{eq.19}
\end{eqnarray} 
This meas that $\widetilde{m}(\infty)$ which determines the total mass of the monopole is not constrained. The matter functions also approach constant asymptotically as
\begin{eqnarray}
 K(\infty) = 0, ~~~H(\infty) = 1.
\label{eq.20}
\end{eqnarray} 
On the other hand, regularity at the origin requires
\begin{eqnarray}
 K(0) = 1, ~~~H(0) = 0,~~~\widetilde{m}(0) = 0.
\label{eq.21}
\end{eqnarray} 

In SU(2) EYMH theory \cite{kn:4}, the gravitating monopole solution emerges smoothly from the flat space 't Hooft-Polyakov monopole ($\alpha = 0$) before becoming a limiting soluton at some critical value of gravitational coupling $\alpha_{\scalebox{.6}{\mbox{c}}}$ and ceases to exist beyond $\alpha_{\scalebox{.6}{\mbox{c}}}$. One generally expects that $\alpha_{\scalebox{.6}{\mbox{c}}}$ should be the maximal value of gravitational coupling, $\alpha_{\scalebox{.6}{\mbox{max}}}$. However, results shows that $\alpha_{\scalebox{.6}{\mbox{max}}}$ does not correspond to the zero of the metric function (in their notation $\mu$) when $\beta = 0$. The tabulated value is $\alpha_{\scalebox{.6}{\mbox{max}}} = 1.403$ corresponds to $\mu_{\scalebox{.6}{\mbox{min}}} = 0.03$.  From $\alpha_{\scalebox{.6}{\mbox{max}}}$ instead the branch of solution bends backward, up to the critical coupling constant $\alpha_{\scalebox{.6}{\mbox{c}}}$ before the zero for $\mu_{\scalebox{.6}{\mbox{min}}}$ is formed and the solution becomes limiting solution (when $\alpha_{\scalebox{.6}{\mbox{c}}} = 1.386, \mu_{\scalebox{.6}{\mbox{min}}} = 8.07 \times 10^{-9}$). In general, the (normalized) mass of gravitating monopole solution decreases with increasing gravitational coupling until the maximal gravitational coupling $\alpha_{\scalebox{.6}{\mbox{max}}} = 1.403$.

Our results in EWS theory are plotted in Fig. \ref{Fig.1}. The metric function $A(x)$ starts to develop a pronounced minimal value ($A_{\scalebox{.6}{\mbox{min}}}$) with increasing $\alpha$ from $\alpha = 0$ up to a maximal value $\alpha_{\scalebox{.6}{\mbox{max}}}$, indicating that gravitating Cho-Maison monopole emerges smoothly from the flat space Cho-Maison monopole. The value of $A_{\scalebox{.6}{\mbox{min}}}$ decreases from one to zero at $\alpha_{\scalebox{.6}{\mbox{max}}}$ where the branch of solution becomes a black hole and ceases to exist. In our solutions, $\alpha_{\scalebox{.6}{\mbox{max}}}$ always corresponds to the lowest value of $A_{\scalebox{.6}{\mbox{min}}}$ ($\alpha_{\scalebox{.6}{\mbox{max}}} = 1.814$ for $A_{\scalebox{.6}{\mbox{min}}} = 3.2992 \times 10^{-6}$). In other words, $\alpha_{\scalebox{.6}{\mbox{max}}} = \alpha_{\scalebox{.6}{\mbox{c}}}$ in EWS theory. We have tried to search for possible lower value of $A_{\scalebox{.6}{\mbox{min}}}$ by considering the coupling constant $\alpha$ bending backwards but the results for $A_{\scalebox{.6}{\mbox{min}}}$ is always higher than the lowest $A_{\scalebox{.6}{\mbox{min}}}$ at $\alpha_{\scalebox{.6}{\mbox{max}}} = 1.814$. These results are tabulated in Table \ref{table.1} (the numerical method used in this paper is different to that of \cite{kn:4}).

Of course the existence of $\alpha_{\scalebox{.6}{\mbox{max}}} > \alpha_{\scalebox{.6}{\mbox{c}}}$ in EYMH theory has been observed for $\beta$ up to 0.7 \cite{kn:4}. It is most profound when $\beta = 0$, where $\alpha_{\scalebox{.6}{\mbox{max}}} = 1.403$ (for $\mu_{\scalebox{.6}{\mbox{min}}} = 0.035$) and $\alpha_{\scalebox{.6}{\mbox{c}}} =  1.386$ (for $\mu_{\scalebox{.6}{\mbox{min}}} = 8.07 \times 10^{-9} $). When $\beta = 0.7$, $\alpha_{\scalebox{.6}{\mbox{max}}} = 1.26027253$ (for $\mu_{\scalebox{.6}{\mbox{min}}} = 0.035$) and $\alpha_{\scalebox{.6}{\mbox{c}}} =  1.2602718$ (for $\mu_{\scalebox{.6}{\mbox{min}}} = 8.07 \times 10^{-9} $). In this case, if large $\beta$ is considered in EYMH theory, one should get $\alpha_{\scalebox{.6}{\mbox{max}}} \approx \alpha_{\scalebox{.6}{\mbox{c}}}$. This has been further confirmed in Ref. \cite{kn:18}. Then question arises if the non-existence of $\alpha_{\scalebox{.6}{\mbox{max}}} > \alpha_{\scalebox{.6}{\mbox{c}}}$ in EWS theory might due to higher value of $\beta$ considered ($\beta = 0.77818833$). 

For the above reason, we first compute the gravitating monopole solutions when $\beta = 0.77818833$ in EYMH theory. Our numerical results shows that even for $\beta = 0.77818833$, the solutions possess phenomena of $\alpha_{\scalebox{.6}{\mbox{max}}} > \alpha_{\scalebox{.6}{\mbox{c}}}$, where $\alpha_{\scalebox{.6}{\mbox{max}}} = 1.2203$ for $A_{\scalebox{.6}{\mbox{min}}} = 1.6388 \times 10^{-5}$ and $\alpha_{\scalebox{.6}{\mbox{c}}} = 1.2123 $ for $ A_{\scalebox{.6}{\mbox{min}}} = 1.2211 \times 10^{-5}$. For the sake of comparison, we also compute gravitating Cho-Maison monopole solution when $\beta \rightarrow 0$. Results again show that in EWS theory $\alpha_{\scalebox{.6}{\mbox{max}}} $ always correspond to lowest value of the metric function $A(x)$. This confirms that the non-existence of $\alpha_{\scalebox{.6}{\mbox{max}}} > \alpha_{\scalebox{.6}{\mbox{c}}}$ is a generic feature of EWS theory. However we are not interested for $\beta \rightarrow 0$ in EWS theory since it is not physical.

In general, our results of gravitating Cho-Maison monopole are quite identical to the gravitating 't Hooft-Polyakov monopole in SU(2) EYMH theory except for the non-existence of `backward bending' in $\alpha$ (or $\alpha_{\scalebox{.6}{\mbox{max}}} > \alpha_{\scalebox{.6}{\mbox{c}}}$). Hence our results can be viewed as having distinctive characteristics compared to that in EYMH theory, though both approach their respective limiting solutions at some specific value of gravitational coupling. Moreover, contrary to gravitating monopole in EYMH theory, results in EWS theory describes a genuine gravitating Cho-Maison monopole turning into a black hole. 

\begin{table}
\caption{Table of $A_{\scalebox{.5}{\mbox{min}}}$ for selected values of $\alpha$ near $\alpha_{\scalebox{.5}{\mbox{max}}}$  for radial excitation (r.e.) and gravitating monopole (g.m.).}
\label{tab:1}       % Give a unique label
\begin{tabular}{lll}
\hline\noalign{\smallskip}
$\alpha$ & $A_{\scalebox{.5}{\mbox{min}}}$(r.e.)  & $A_{\scalebox{.5}{\mbox{min}}}$(g.m.)  \\ 
\noalign{\smallskip}\hline\noalign{\smallskip}
1.53 & $2.5783 \times 10^{-4}$ & 0.1849 \\ 
1.54 & $1.8504 \times 10^{-4}$ & 0.1771 \\ 
1.55 & $1.2296 \times 10^{-4}$ & 0.1694 \\ 
1.56 & $7.2356 \times 10^{-5}$ & 0.1617 \\ 
1.57 & $1.2173 \times 10^{-5}$& 0.1540 \\ 
1.58 & $6.7480 \times 10^{-6}$  & 0.1464 \\ 
1.584 & $5.4718 \times 10^{-6}$  & 0.1434 \\ 
1.76 & - & 0.0243 \\ 
1.77 & - & 0.0188\\ 
1.78 & - & 0.0137 \\ 
1.79 & - & 0.0089 \\ 
1.80 & - & 0.0045 \\ 
1.81 & - & $8.2190 \times 10^{-4}$ \\ 
1.814 & - & $3.2992 \times 10^{-6}$  \\ 
\noalign{\smallskip}\hline
\end{tabular}
\label{table.1}
\end{table}

\section{Radially Excited Monopole}
\label{sec:4}

\begin{figure}[!b]
	\centering
	\hskip0in
	 \includegraphics[width=3.3in]{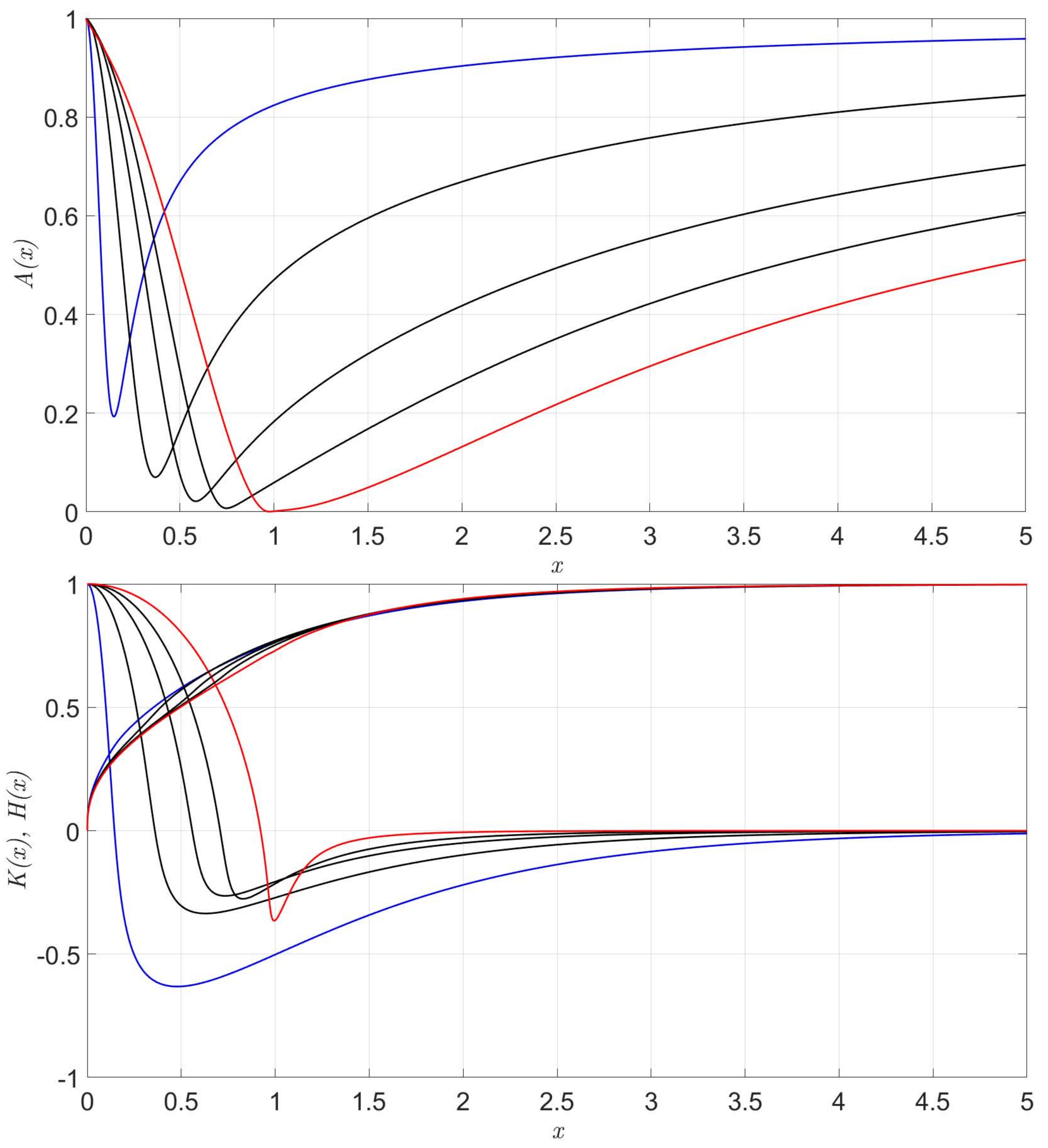} 
	\caption{Functions of $A(x)$ , $K(x)$ and $H(x)$ versus $\alpha$ of the radial excitation for $\alpha =0.2 $ (blue), $0.6, 1.0, 1.25$ and $1.5$ (red), for physical $\beta$ and Weinberg angle.}
 \label{Fig.2}
\end{figure}

For EYMH theory, besides the branch of fundamental gravitating monopole solution, there exist branches of radially excited monopole solutions. While the gauge field function of the fundamental monopole solution decreases monotonically to zero, this is not the case for radially excited solutions. In general, gauge field function of the $n$-th excited monopole solutions develop $n$ minimum node before tending to zero at spatial infinity. Similar to fundamental monopole solution, radially excited monopole solutions also exist below some maximal value of the gravitational constant $\alpha$. However, they have no flat space counterpart as $\alpha \rightarrow 0$, but tends to the Bartnik-Mckinnon solution of EYM theory .

In EWS theory, we also observed similar radially excited monopole solution, as shown in Fig. \ref{Fig.2}. The gauge field function $K(x)$ of the (1st) radially excited solution does not decrease monotonically to zero, it develops a minimum node before approaches zero at spatial infinity. These radial excitations only exist below some maximal value of the gravitational constant, $\alpha_{\scalebox{.6}{\mbox{max}}} = 1.584$. They similarly do not have flat space counterpart as $\alpha \rightarrow 0$. Following the gravitating monopole in previous section, we also tabulate the value of $A_{\scalebox{.6}{\mbox{min}}}$ for values of $\alpha$ near $\alpha_{\scalebox{.5}{\mbox{max}}}$  in Table 1. We again observed no `backward-bending' of the solution branch as reported in Ref. \cite{kn:4}.

\begin{figure}[!b]
	\centering
	\hskip0in
	 \includegraphics[width=3.3in]{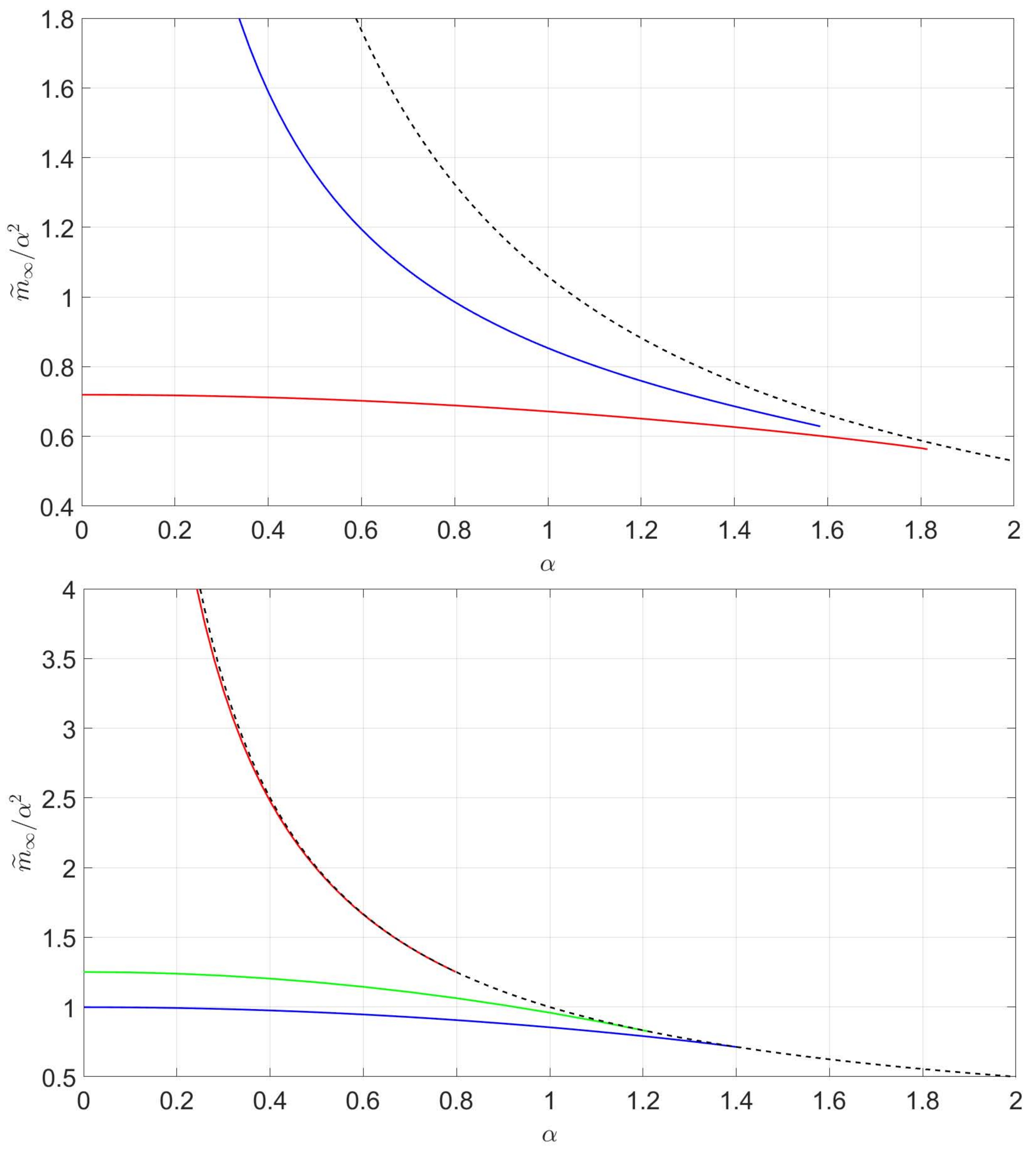} 
	\caption{Top: Plot of $\widetilde{m}_{\infty}/\alpha^2$ versus $\alpha$ for gravitating monopole (blue), radial excitation (red) and the extremal RN solution (dashed line) of the EWS theory. Bottom: Plot of $\widetilde{m}_{\infty}/\alpha^2$ versus $\alpha$ for gravitating monopole (blue), radial excitation (red) and the extremal RN solution (dashed line) of the EYMH theory when $\beta = 0$. Green line represents the gravitating monopole when $\beta = 0.77818833$.}
\label{Fig.3}
\end{figure}

Following Ref. \cite{kn:17}, we plot the normalized mass $\widetilde{m}_{\infty}/\alpha^2$ versus $\alpha$ for radial excitation in Fig. \ref{Fig.3} (included in the same graph is the corresponding plot for fundamental gravitating monopole). As expected, the radial excitation branch possesses higher normalized mass than that of the fundamental gravitating monopole.  In the limit of $\alpha \rightarrow 0$, the normalized mass of the fundamental gravitating monopole converges to a finite value (0.7197), whereas that of radial excitation diverges to infinity. This coincides with the statement that radial excitation does not have flat space counterpart. Both massess of the fundamental monopole and radial excitation decrease with $\alpha$, but mass of the gravitating monopole decreases monotonically while there is an inverse $\alpha$ fall-off for radial excitation.

At $\alpha_{\scalebox{.6}{\mbox{max}}}$, the radial excitation (as well as the fundamental monopole solutions) reaches their limiting functions but do not bifurcate with the branch of extremal RN solution. This is different from the results reported in Ref. \cite{kn:17}, where at $\alpha_{\scalebox{.6}{\mbox{max}}}$ the fundamental monopole approaches its limiting solution and  bifurcate with the branch of extremal RN solution (bottom plot of Fig. \ref{Fig.3}). The non-bifurcation originates from Eq. (\ref{eq.16}). Recall that in EYMH theory, the metric function of the embbed RN solution with magnetic charge $P$ reads 
\begin{eqnarray}
A = 1 - \frac{\widetilde{m}_{\infty}}{x} + \frac{\alpha^2}{x^2} P^2.
\label{eq.22}
\end{eqnarray}
Eq. (\ref{eq.16}) gives metric function of the embbed RN solution of EWS theory as
\begin{eqnarray}
A = 1 - \frac{\widetilde{m}_{\infty}}{x} + \frac{\alpha^2}{x^2} \frac{1}{4} \left(  1 + \frac{\epsilon}{\omega^2}  \right).
\label{eq.23}
\end{eqnarray}
From Eq. (\ref{eq.23}), evaluating the third term by considering $\epsilon = 1$ and $\omega = 0.53574546$, we find that the magnetic charge of embeded RN solution is 1.0588, which is slightly higher than one and this  contributes to the non-bifurcation (note that the gravitating monopole or radially excited monopole has unit magnetic charge). 

Hence in general our results of radially excited Cho-Maison monopole are quite identical to the radially excited monopole in SU(2) EYMH theory. There are of course some key differences. First, radial exictation (as well as gravitating monopole) does not bifurcate with the branch of extremal RN solution at $\alpha_{\scalebox{.6}{\mbox{max}}}$. We expect that a different (more realistic) form of $\epsilon$ will contribute to the bifurcation, but that remains to be answered in future investigation. Second, for a given $\alpha$, the mass of radial excitation (or gravitating monopole) in EWS theory is always lower than the mass of their counterpart in EYMH theory. This is evident from Fig. \ref{Fig.3}.

\section{Black Hole Solutions}
\label{sec:5}

\begin{figure}[!b]
	\centering
	\hskip0in
	 \includegraphics[width=3.3in]{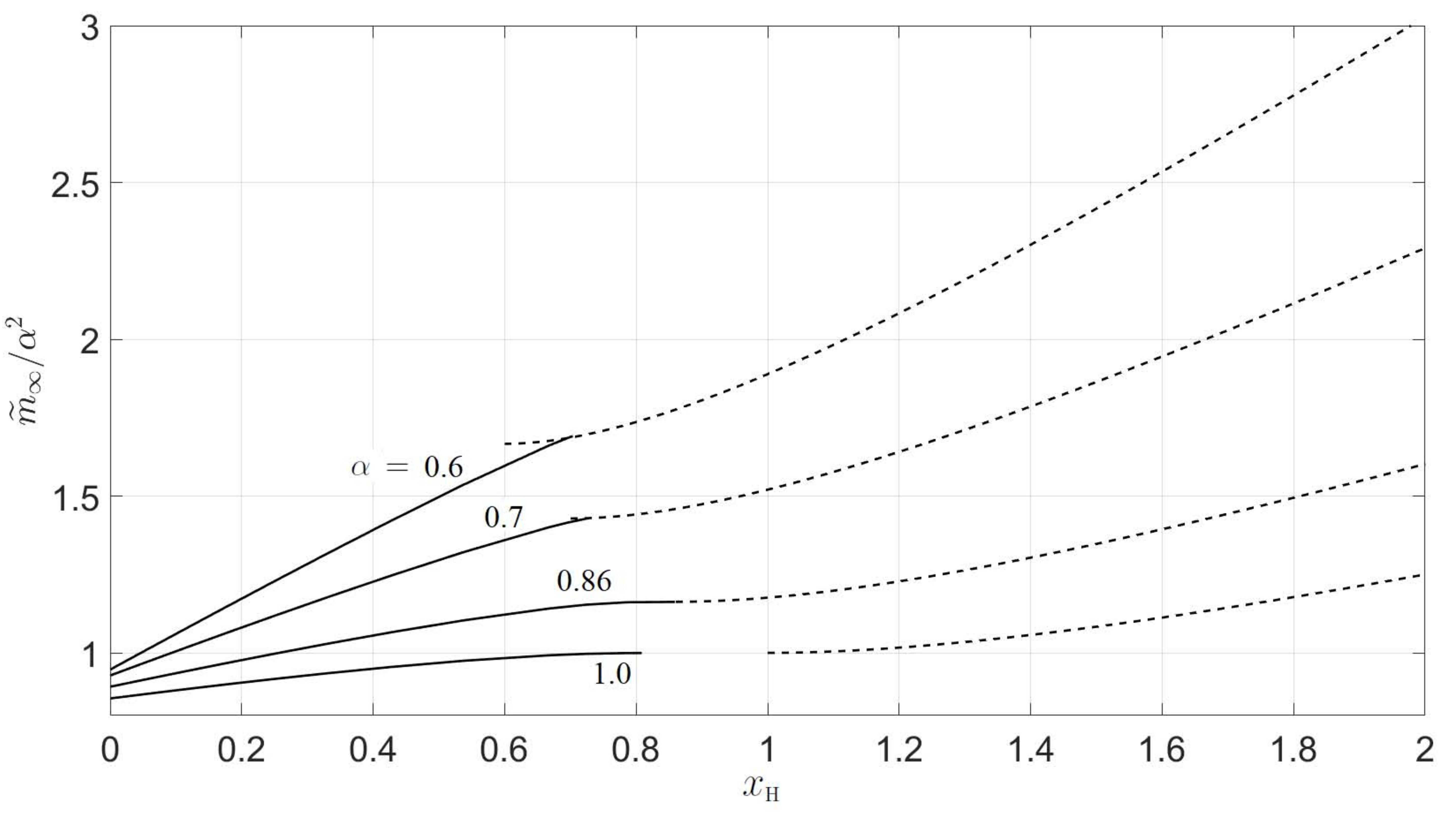} 
	\caption{The (normalized) mass of the EYMH black hole solutions $\widetilde{m}_{\infty} / \alpha^2$ as a function of horizon radius $x_{\scalebox{.6}{\mbox{H}}}$ for $\alpha = 0.6, 0.7, 0.86$ and $1.0$, together with the corresponding RN solutions with unit magnetic charge (dashed lines).}
 \label{Fig.4}
\end{figure}

In SU(2) EMYH theory, there exist special kind of non-Abelian black hole solutions which is different from the embedded RN black hole solutions. These black hole solutions emerge from the globally regular monopole solution when a finite regular event horizon is imposed. Characterized as `black hole within magnetic monopole', these solutions provide counter-examples to the `no hair' conjecture. With increasing horizon radius $x_{\scalebox{.6}{\mbox{H}}}$, depending on the value of gravitational coupling $\alpha$, these non-Abelian black hole either merges with Abelian RN black hole at some critical horizon radius (for $0 < \alpha < \frac{1}{2} \sqrt{3}$), or ceases to exist at some maximal value of $x_{\scalebox{.6}{\mbox{H}}}$ when a second zero of the metric function is formed (for $\frac{1}{2} \sqrt{3} < \alpha < \alpha_{\scalebox{.6}{\mbox{max}}}$, where $\alpha_{\scalebox{.6}{\mbox{max}}} = 1.403$). These behaviours of SU(2) EYMH black hole solutions are shown in Fig. \ref{Fig.4}.

\begin{figure}[!b]
	\centering
	\hskip0in
	 \includegraphics[width=3.3in]{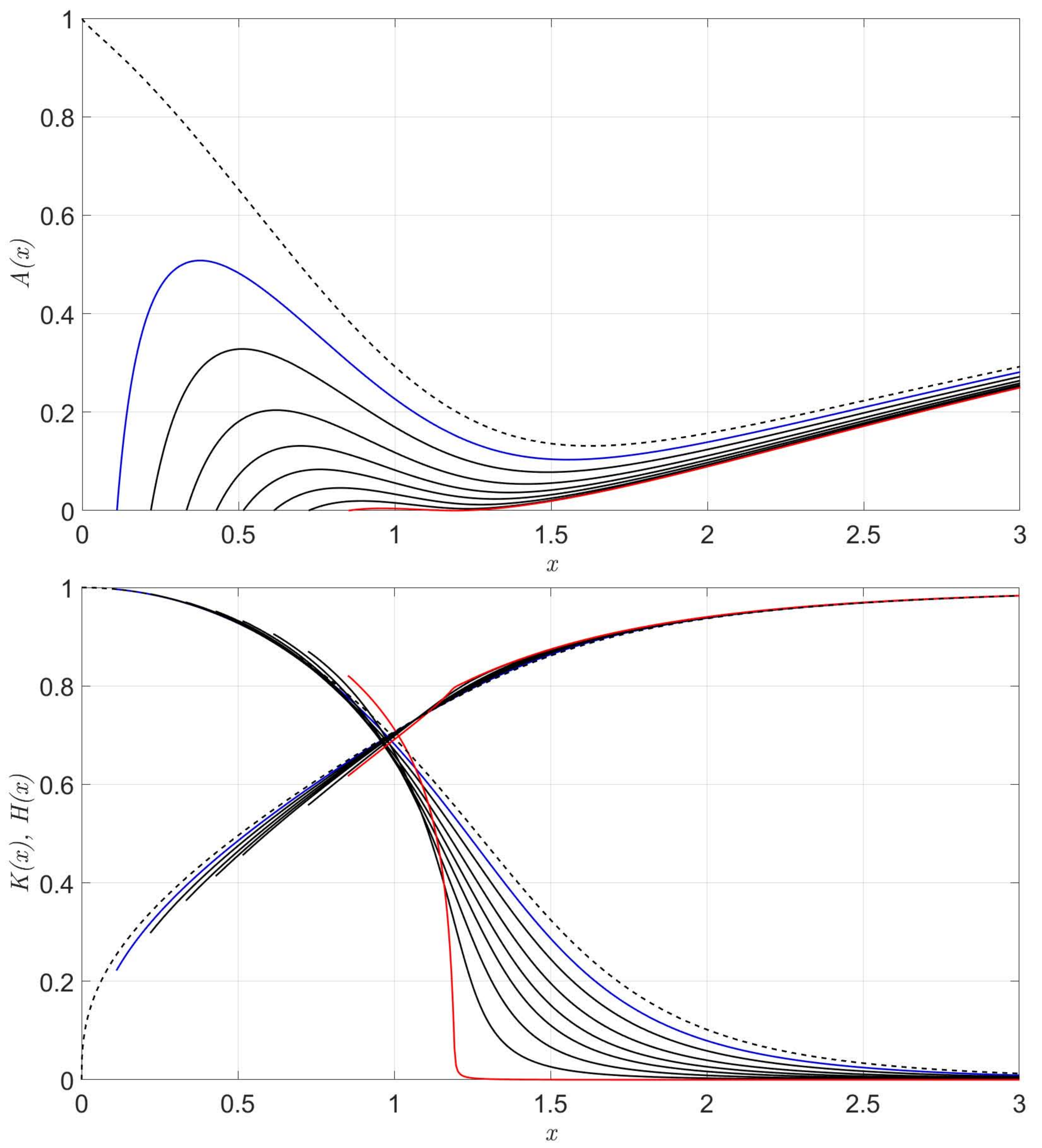} 
	\caption{Black hole in electroweak monopole solutions with $\alpha = 1.6$, physical $\beta$ and Weinberg angle, plotted as a function of $x$ for $x_{\scalebox{.6}{\mbox{H}}} = 0.1111$ (blue), $0.2195, 0.3333, 0.4286, 0.5152, 0.6129, 0.7241$ and $0.8519$ (red). Dashed lines indicate regular gravitating solutions.}
 \label{Fig.5}
\end{figure}

We now consider such non-Abelian black hole solutions in EWS theory. Consider again asymptotic flatness, they satisfy the same boundary conditions at spatial infinity as the globally regular solutions, Eqs. (\ref{eq.19}) and (\ref{eq.20}). The existence of a regular event horizon requires
\begin{eqnarray}
\widetilde{m} \left(  x_{\scalebox{.6}{\mbox{H}}} \right) = \frac{x_{\scalebox{.6}{\mbox{H}}}}{2},~~~N \left(  x_{\scalebox{.6}{\mbox{H}}} \right) < \infty.
\label{eq.24}
\end{eqnarray}
The matter functions must also satisfy
\begin{eqnarray}
\left. \frac{dA}{dx} \frac{dK}{dx}  \right|_{x_{\scalebox{.6}{\mbox{H}}}} = \left.  K \left(   H^2  - \frac{1 - K^2}{x^2} \right)    \right|_{x_{\scalebox{.6}{\mbox{H}}}},
\label{eq.25}
\end{eqnarray}
and
\begin{eqnarray}
&& \left. \frac{dA}{dx} \frac{dH}{dx}  \right|_{x_{\scalebox{.6}{\mbox{H}}}}= \nonumber\\
&& \left.  H \left( \frac{K^2}{2 x^2} + 2 \beta^2 \left( H^2 - 1 \right) + \frac{1}{2 \omega^2 x^4 H} \frac{d \epsilon}{d H}  \right)  \right|_{x_{\scalebox{.6}{\mbox{H}}}}.
\label{eq.26}
\end{eqnarray}
 
\begin{figure}[!b]
	\centering
	\hskip0in
	 \includegraphics[width=3.3in]{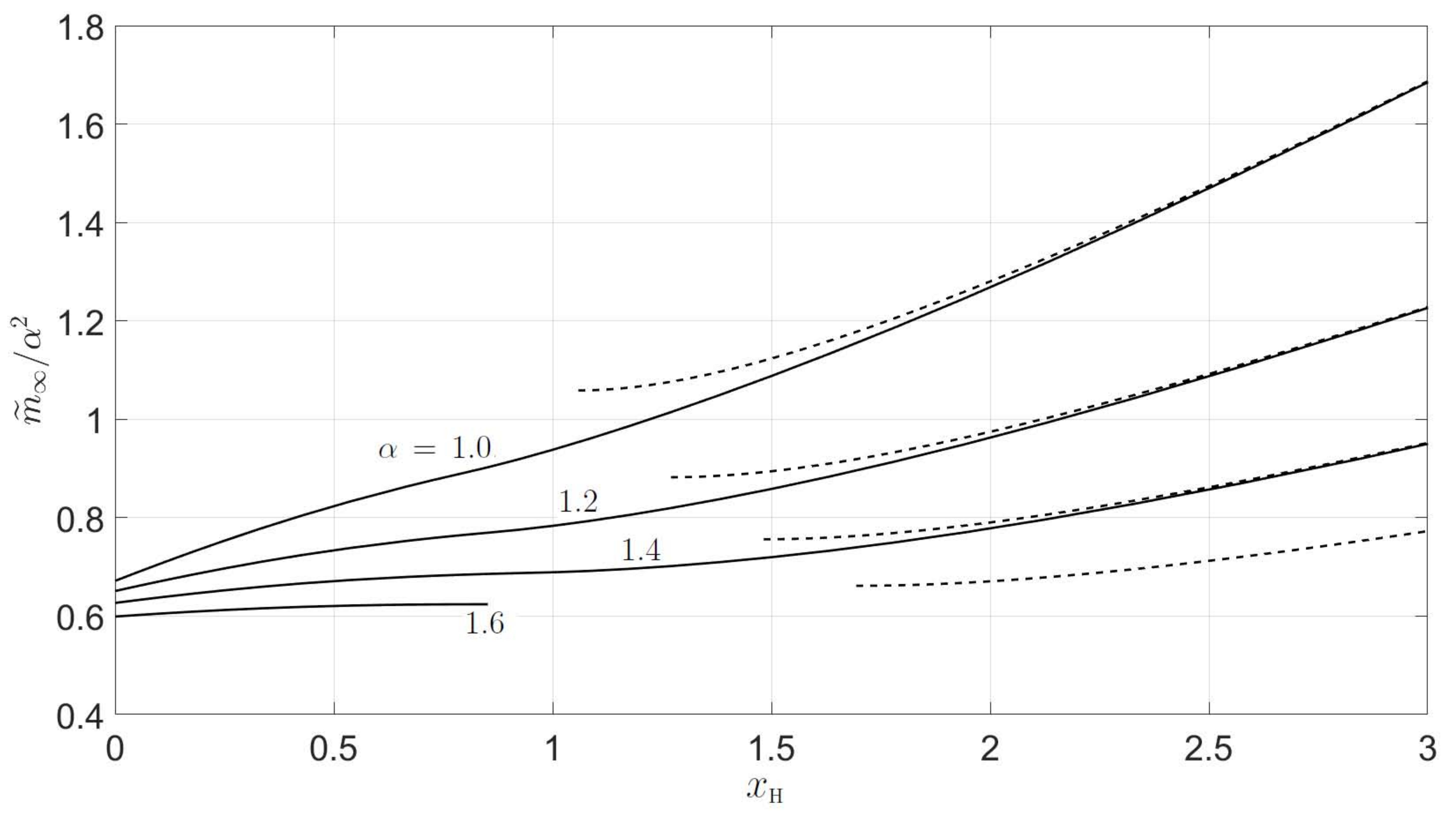} 
	\caption{The (normalized) mass of the EWS black hole solutions $\widetilde{m}_{\infty} / \alpha^2$ as a function of the horizon radius $x_{\scalebox{.6}{\mbox{H}}}$ for the values of coupling constant $\alpha = 1.0, 1.2, 1.4$ and $1.6$, together with the corresponding RN solutions with unit magnetic charge (dashed lines).}
 \label{Fig.6}
\end{figure}

In particular, for a given coupling constant $\alpha$, black hole solutions corresponding to the fundamental mono-pole branch emerge from globally regular solution in the limit $x_{\scalebox{.6}{\mbox{H}}} \rightarrow 0$ and persist up with increasing horizon radius. We first consider the case of relatively large $\alpha$ ($\alpha = 1.6$). With increasing horizon radius, limiting solution is reached at a maximal value of horizon radius ($x_{\scalebox{.6}{\mbox{H}}} = 0.8519$), where a second zero of $A(x)$ is formed, Fig. \ref{Fig.5}. For smaller values of $\alpha$, black hole solutions do not reach limiting solutions but merge with the corresponding non-extremal RN solutions. This behaviour which is reminiscent of SU(2) EYMH theory can be understood clearer from Fig. \ref{Fig.6}, which shows the black hole solutions emerge from globally regular monopole solutions in the limit of $x_{\scalebox{.6}{\mbox{H}}} \rightarrow 0$ and persist up with increasing $x_{\scalebox{.6}{\mbox{H}}}$. For $0 < \alpha < 1.576$, the black hole solutions slowly converge to the corresponding non-extremal RN solutions at large horizon radius. For $1.576 < \alpha < \alpha_{\scalebox{.6}{\mbox{max}}}$ where $\alpha_{\scalebox{.6}{\mbox{max}}} = 1.814$, they however become limiting solution at maximal value of horizon radius. 

\begin{figure}[!b]
	\centering
	\hskip0in
	 \includegraphics[width=3.3in]{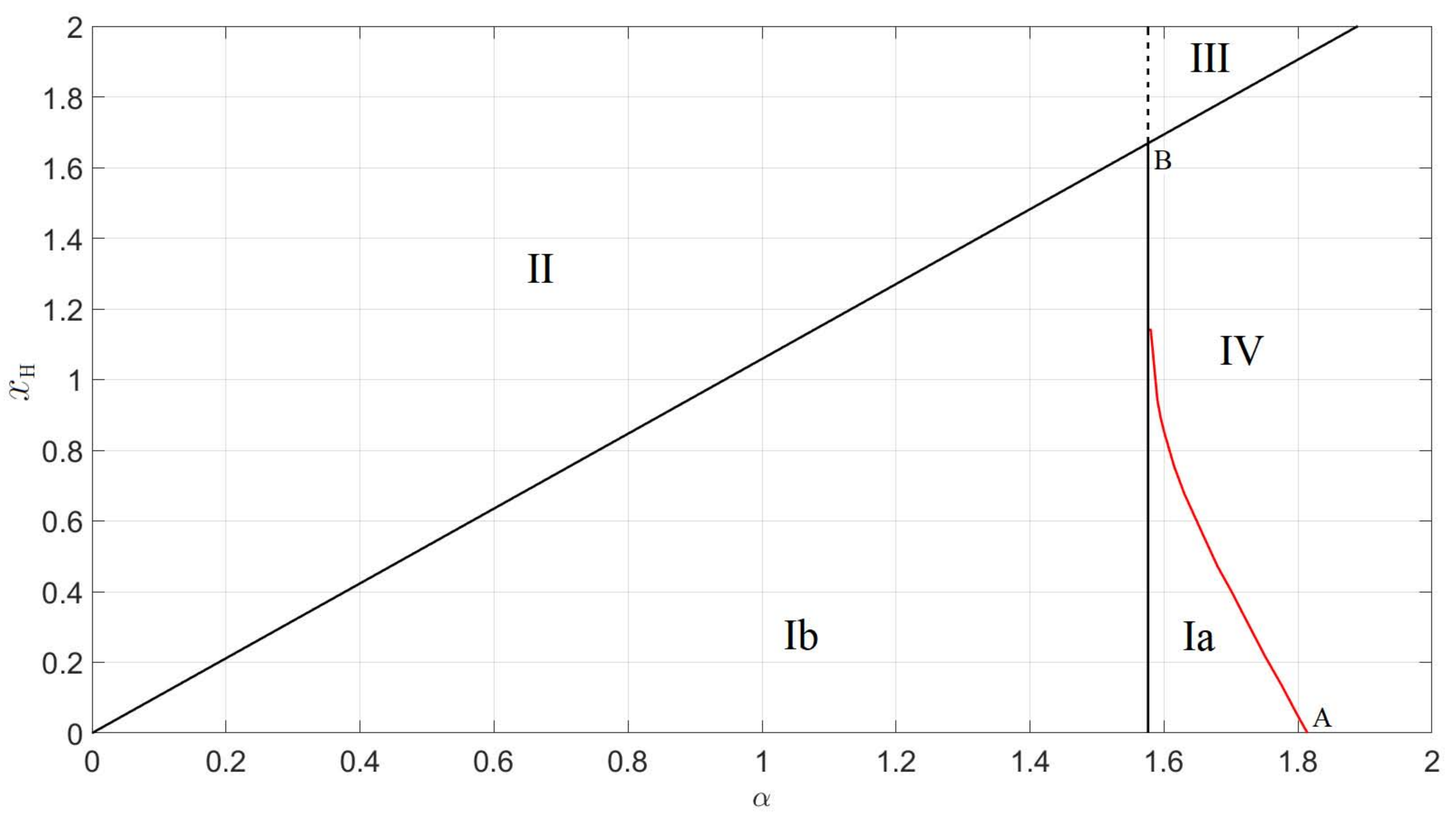} 
	\caption{`Phase diagram' of black hole solution in Einstein-Weinberg-Salam theory for physical $\beta$ and Weinberg angle.}
 \label{Fig.7}
\end{figure}

To better illustrate the solutions, following Ref. \cite{kn:4}, we also present the `phase diagram' of black hole solution in EWS theory for physical $\beta$ and Weinberg angle in Fig. \ref{Fig.7}. Non-Abelian black hole exist in regions of the ($\alpha, x_{\scalebox{.6}{\mbox{H}}}$) plane denoted by I and II. RN black holes as given by Eq. (\ref{eq.16}), exist in regions II and III, whereas in region IV there are no non-Abelian black hole solutions.  The boundary $x_{\scalebox{.6}{\mbox{H}}} = 0$ of region I corresponds to the regular gravitating solutions. The cut-off point of $\alpha = 1.576$ as mentioned above is seen as separating region I into Ia and Ib. Approaching the curve AB in region Ia, the solutions develop double zero in $A(x)$ and they become limiting solutions. In region Ib, the non-Abelian solutions extend into region II and slowly merge into the RN solution with increasing $x_{\scalebox{.6}{\mbox{H}}}$.

Hence the `black hole in Cho-Maison monopole' of EWS theory again have identical features as the `black hole in monopole' of SU(2) EYMH theory \cite{kn:4}, \cite{kn:16}. There are however some key differences:\\
\noindent 1. First (for small $\alpha$) the EWS black hole solutions do not merge with the corresponding RN solution at critical horizon radius, but only converges towards them slowly with increasing horizon radius. The EYMH black hole solutions merge with the non-extremal RN solutions at critical value of the horizon radius  \cite{kn:17}. \\
\noindent 2. Second, for region of ($\alpha, x_{\scalebox{.6}{\mbox{H}}}$) plane where non-Abelian black hole and RN black holes coexist, the mass of non-Abelian black hole solution is always lower than the mass of RN solution ($ m_{\scalebox{.6}{\mbox{n.a.}}} / m_{\scalebox{.6}{\mbox{RN}}} \leq 1$). The case in EYMH theory is however more complicated. The region of coexistence for non-Abelian black hole and RN solution exists for $ 0 < \alpha < 0.77$. Here the mass of non-Abelian black hole is smaller than the mass of RN solution ($m_{\scalebox{.6}{\mbox{n.a.}}} / m_{\scalebox{.6}{\mbox{RN}}} < 1$), but there also exists small region where $m_{\scalebox{.6}{\mbox{n.a.}}} / m_{\scalebox{.6}{\mbox{RN}}} > 1$. Then for $ 0.77 < \alpha < 0.866$, non-Abelian black hole solution joints smoothly with the RN solution at critical value of horizon radius. For $0.866 < \alpha <  1.403$, non-Abelian black hole solution reach limiting solution at maximal value of horizon radius, and does not bifurcate with the RN solution. \\
\noindent 3. Third, for a given horizon radius $x_{\scalebox{.6}{\mbox{H}}} $, the mass of black hole in EWS theory always has a lower value than its counterpart in EYMH theory.

\section{Conclusions}
\label{sec:6}

We have studied numerical solutions of the Einstein-Weinberg-Salam theory corresponding to: (1) fundamental gravitating electroweak monopole; (2) radially excited electroweak monopole; and (3) non-Abelian magnetically charged black hole. 

The fundamental monopole solution emerges from the corresponding flat space monopole solution and extends smoothly up to a maximal value of the gravitational coupling constant $\alpha_{\scalebox{.5}{\mbox{max}}}$, before they collapse into a black hole. Besides the fundamental monopole branch, there exist branches of radially excited monopole solution, which also exist up to a maximal value of $\alpha$. However there are no flat space counterpart in the limit of $\alpha \rightarrow 0$ for the radially excited solution.

Both the normalized mass of fundamental monopole and radial excitation decrease with $\alpha$, with radial excitation branch possesses higher mass than that of the fundamental gravitating monopole. In the limit of $\alpha \rightarrow 0$, the normalized mass of the fundamental monopole branch converges to a finite value (0.7197), indicating the ADM mass of approximately 6.821 TeV. On the other hand the normalized mass of radial excitation diverges to infinity as $\alpha \rightarrow 0$, indicating that the radial excitation does not have flat space counterpart. We summarize the numerical estimate of ADM mass for gravitating monopole and radial excitation for selected values of $\alpha$ in Table \ref{table.2}.

\begin{table}
\caption{The numerical estimate of ADM mass for gravitating monopole (g.m.) and radial excitation (r.e.) with $\epsilon = \left( H/H_0\right)^8$ and physical value of $\beta$.}
\label{tab:1}       % Give a unique label
\begin{tabular}{lll}
\hline\noalign{\smallskip}
$\alpha$ & ADM mass (r.e.) & ADM mass (g.m.)  \\
\noalign{\smallskip}\hline\noalign{\smallskip}
0 & $\infty$ &  6.821 TeV \\ 
0.20 & 25.555 TeV &  6.802 TeV \\ 
0.40 & 15.058 TeV &  6.747 TeV \\ 
0.60 & 11.320 TeV & 6.656 TeV \\ 
0.80 & 9.344 TeV &  6.529 TeV \\ 
1.00 & 8.089 TeV &  6.367 TeV \\ 
1.20 & 7.198 TeV &  6.171 TeV \\ 
1.40 & 6.506 TeV &  5.942 TeV \\ 
1.584 & black hole  &  5.702 TeV \\ 
1.80 & - &  5.369 TeV \\ 
1.814 & - & black hole \\
\noalign{\smallskip}\hline
\end{tabular}
\label{table.2}
\end{table}

For the `black hole in electroweak monopole', black hole solutions corresponding to fundamental monopole branch emerge from globally regular solution in the limit $x_{\scalebox{.5}{\mbox{H}}} \rightarrow 0$ and persist up differently (depending on coupling constant $\alpha$) with increasing horizon radius. For a relatively large $\alpha$ ($1.576 < \alpha < 1.814$), limiting solution is reached at a maximal value of horizon radius, e.g. $x_{\scalebox{.5}{\mbox{H}}}(\mbox{max})= 0.8519$ for $\alpha = 1.6$. However for smaller values of $\alpha$ ($0 < \alpha < 1.576$), black hole solutions do not reach limiting solutions but slowly merge into the corresponding non-extremal RN solutions at large horizon radius. 

Despite mostly identical, our results in Einstein-Weinberg-Salam theory have some key differences compared to that of EYMH theory: (1) Our solutions of gravitating monopole and radial excitation do not exhibit the phenomena of `backward-bending' in coupling constant $\alpha$ as observed in Refs. \cite{kn:4,kn:17}. The maximal value $\alpha_{\scalebox{.6}{\mbox{max}}}$ always correspond to the lowest value of the metric function $A$ and the solutions become limiting solutions at $\alpha_{\scalebox{.6}{\mbox{max}}}$. (2) At $\alpha_{\scalebox{.6}{\mbox{max}}}$, both the (normalized) mass of gravitating monopole and radial excitation reaches minimum value but do not bifurcate with the branch of extremal RN solution respectively. (3) The non-Abelian black hole solution converges slowly into the corresponding non-extremal RN solutions with large horizon radius. At a given horizon radius where the non-Abelian black hole and RN solution coexist, the mass of non-Abelian black hole is always lower than the mass of RN solution. (4) At a given $\alpha$, the mass of gravitating monopole (or radial excitation) in EWS theory always has lower value than its counterpart in EYMH theory. Similarly, at a given horizon radius $x_{\scalebox{.6}{\mbox{H}}} $, the mass of black hole in EWS theory is also lower than the mass of its counterpart in EYMH theory. This suggests that the configurations in EWS theory are more stable.

In SU(2) EYMH theory, gravitating monopoles and magnetically charged black holes can be generalized to gravitating dyons and dyonic black holes \cite{kn:17}. Hence our results here clearly have the dyonic generalization simply by switching on the time component of the gauge potential. We will discuss these findings in a separate paper.

%\begin{acknowledgements}
%\end{acknowledgements}



\end{document}